\documentclass[letterpaper]{article}
\usepackage{mume}
\usepackage{times}
\usepackage{helvet}
\usepackage{courier}
\usepackage{graphicx}
\usepackage{xparse}
\usepackage{url}
\usepackage[dvipsnames]{xcolor}
\NewDocumentCommand{\codeword}{v}{%
\texttt{\textcolor{black}{#1}}%
}
\begin{document}
	%
	\title{Apollo: An Interactive Environment for Generating Symbolic Musical Phrases using Corpus-based Style Imitation}
	\author{Renaud Bougueng Tchemeube, Jeff Ens, Philippe Pasquier \\
		\{rbouguen, jeffe, ppasquier\} at sfu dot ca \\ Simon Fraser University
	}
	\maketitle
	\begin{abstract}
		\begin{quote}
		     With the recent developments in machine intelligence and web technologies, new generative music systems are being explored for assisted composition using machine learning techniques on the web. Such systems are built for various tasks such as melodic, harmonic or rhythm generation, music interpolation, continuation and style imitation. In this paper, we introduce Apollo, an interactive music application for generating symbolic phrases of conventional western music using corpus-based style imitation techniques. In addition to enabling the construction and management of symbolic musical corpora, the system makes it possible for music artists and researchers to generate new musical phrases in the style of the proposed corpus. The system is available as a desktop application. The generated symbolic music materials, encoded in the MIDI format, can be exported or streamed for various purposes including using them as seed material for musical projects. We present the system design, implementation details, discuss and conclude with future work for the system.
		\end{quote}
	\end{abstract}
	
	\section{Introduction}
	
	Computer-assisted composition (CAC) is an area of computer music that explores technological formalisms for digitally assisting composers in the process of ideating and developing musical pieces \cite{assayag1998computer}. This implies that the computer makes compositional decisions using algorithmic processes \cite{papadopoulos1999ai}. Generative music systems are an example of such systems where algorithms are used to generate original musical pieces or to assist the composer in doing so. These algorithms fall within the umbrella of musical metacreation, a field concerned with endowing computers with the partial or complete automation of musical tasks \cite{pasquier2016introduction}. Many such algorithms are developed based on artificial intelligence (AI) and machine learning (ML) techniques \cite{fernandez2013ai}.
	In many of these systems, the composer has limited control over the shaping of the algorithm's behavior. Algorithms are often pre-trained and presented as black boxes to the users. The user usually only gets to tweak some generative parameters to make music. In contrast, the approach of interactive machine learning (IML) consists in exposing access to the other aspects of the ML process such as corpus preparation and training activities in a way that involves a feedback loop \cite{fails2003interactive}.  
	CAC systems are deployed on various platforms, from desktop applications to plugins for music authoring software and web applications running in the browser.
	With advances in computer power for ML and web technologies for multimedia support, web-based generative systems for assisted-composition are being investigated \cite{waite2016project,thio2019minimal}. However, music generative systems  that support corpus manipulation or the ability to change the underlying ML model are lacking. 
	This motivates the development of the Apollo system, an interactive music environment for artists and researchers to explore generative music using algorithms of style imitation in a creative workflow supported by the IML process. A corpus manager is available to the user to prepare and manage music corpora prior to training style imitation algorithms. The user can select an algorithm to be trained. The user can then generate new musical phrases by tweaking generation parameters available via the graphical user interface (GUI) of the system. Our system differentiates itself from other existing systems by engaging with a design approach that focuses on direct open control over the ML process and continuous end-user interaction with corpus manipulation, model selection, training, and control of generation parameters. A demo of the system can be accessed at \textcolor{blue}{\textbf{http://metacreation.net/apollo/}}. 
	
	In the next section, we cover background on computer-assisted music composition, musical metacreation and interactive machine learning. We then review work on related generative music systems, present our system and its implementation details, discuss and conclude with future work.

	\section{Background}
    \subsection{Computer-assisted Music Composition}
    Computer-aided composition (CAC) is a subfield of Computer Music which is concerned with the development of computer technologies and systems that can support the tasks of the compositional process; mainly exploring, developing and rendering musical ideas.
	Such technologies can assist the composition process by supporting the writing and editing of symbolic representations of music, and the ability to render them to audio signals. CAC systems exist in many varieties and are investigated for a broad range of objectives. They can be developed to handle symbolic or acoustic representations of music. Digital symbolic representations are typically encoded using the Musical Instrument Digital Interface (MIDI) format. Communication protocols such as Open Sound Control (OSC) allows for ubiquitous interoperability between two or more digital music systems on a computer network \cite{wright1997open}.
	
    Herremans et al. propose a functional taxonomy of music generation systems \cite{herremans2017functional}. The taxonomy is organized as a concept map centered around music composition as a functional task consisting of melody, harmony, rhythm and timbre. Generative music systems are organized given the degree to which they support a \textit{narrative} approach, \textit{interactive composing}, or \textit{difficulty of performing} (of operating the system). Given those functional themes, the authors further break down the main algorithmic techniques used to approach them (Markov models, factor oracles, neural networks, etc).
    The Apollo system as a compositional tool fits within the functional themes of interaction with support for rhythmic, melodic and harmonic generation.
	
	Algorithmic composition approaches have been explored to support the creative output of the compositional process. These algorithms are pieces of code capable of generating new musical compositions. They can be situated as part of the domain of metacreation.
	
	\subsection{Musical Metacreation and Style Imitation}
	Metacreation is a field of research that is concerned with the partial or complete automation of creative tasks \cite{pasquier2016introduction}. It differs from traditional artificial intelligence (AI) in that the problems it addresses do not have an optimal solution. Musical metacreation is a subfield of metacreation which addresses music-related creative tasks. Those tasks are usually around music composition, interpretation, improvisation and accompaniment. The generative music systems can be online or offline. 
	
	Another important task is that of style imitation. Style imitation is a task which consists in generating new instances that will be labeled to belong to a style S by an unbiased observer given a corpus C of music depicting that style \cite{pasquier2016introduction}. A concrete example of such task is the problem of  generating of chorales in the style of Bach. This particular problem has been extensively explored by the community \cite{liang2016bachbot,hadjeres2017deepbach}.
    The task of style imitation is thus a generative problem whose settings are very similar to that of unsupervised ML problems \cite{dubnov2003using,assayag2001automatic}. As a result, achieving generative capabilities often relies on the usage of statistical modeling techniques from machine learning \cite{briot2017deep}.
    There are many algorithms that imitate the style of the corpus on which they are trained. For our work, we want to be able to choose and train algorithms dynamically by following the interactive machine learning.
	
    \subsection{Interactive Machine Learning for Music}
    Interactive machine learning refers to the idea that, given the complexity and effort involved in machine learning activities such as experimenting with data processing or tweaking training parameters of algorithms, the machine learning process should be an interactive one \cite{fails2003interactive}. It is a way of human-computer interaction where the user engages in a feedback loop process to bring  the system to the desired learning behavior.
    
    The machine learning process uses a corpus of data examples to train algorithms. The corpus of data is fed to the learning algorithm for training, with some parametrization usually involved. Following training, the algorithm can be employed for the intended purposes. The expectation that a machine learning model can be trained only once is often untrue. Furthermore, training machine learning models effectively has been described to be more an art than a science \cite{domingos2012few}. It is also a complex discipline that requires strong technical expertise and the ability to write code. As a result, designing new interfaces for computer-assisted music composition using machine learning becomes a challenge \cite{poupyrev2001new}. This is particularly true when thinking of machine learning algorithms as creative user tools for musical expression \cite{fiebrink2016machine}. This opens up the opportunity to present the process of machine learning as a creative process using music generative systems to the user; effectively making it part of his creative musical process.
    This is a valuable approach to interfacing generative systems given that ``interactivity is an essential performative aspect of the musical discovery  process" \cite{agostini2013real}.
	
	\section{Related Work}
	In this section, we review interactive generative music systems relevant to our work. We focus on those using machine learning to generate music for style imitation.

	
    
    The Style Machine \cite{anderson2013generative}\cite{eigenfeldt2013evolving} is an interactive corpus-based system with the ability to generate partial or complete musical pieces given a corpus of the style of electronic dance music (EDM). This includes complete musical arrangements and the ability to re-generate parts for a given instrument. The generation is controlled by three parameters: complexity, density and length of the generated pieces.

   The open source Magenta project \cite{waite2016project} offers a library for music generation in both Python and JavaScript languages. The library hosts a growing collection of ready-to-use deep learning algorithms for music. The user programmatically instantiates a model and can use it to generate new musical content. Although, many web applications have been developed that uses these models (Latent Loops \footnote{https://magenta.tensorflow.org/composing-palettes}, Multitrack MusicVAE \footnote{https://magenta.tensorflow.org/multitrack}), including systems with support for standalone installation and plugins for Digital Audio Workstations (DAW) such as Ableton (Magenta Studio \footnote{https://magenta.tensorflow.org/studio}), they do not allow the model to be trained by the user. The user might feed musical data to the models only in the context of outputting new music e.g. feed data as seed for generation. As a result, user interactivity with the machine learning process is limited.
	
	Thio et al. present a minimal template for implementing demonstrations of deep learning music models on the web \cite{thio2019minimal}. The template does not provide support for the user to manage music corpora and feed them to the models. This means that the learning algorithms are pre-trained and are only accessible to the user for the generative activities. 

    ManuScore is an interactive notation-based music composition application using a cognitively-inspired music learning and generation system called MusicCog \cite{maxwell2014generative}. The system is intended to allow composers to experiment with ``interactive, generative, object-oriented  composition", with minimal disruption to their existing  musical language. MusicCog is a modular cognitive architecture implementing functions of music perception, working memory, long-term memory, and music composition. It is capable of performing polyphonic voice-separation, melodic segmentation, chunking and hierarchical sequence learning. ManuScore supports  using musical gestures for note entry on  staff and transfer of staff data between a source and a target staff. The data supported for transfer are 1) pitch and rhythmic contours, 2) pitch, harmonic and rhythmic grids using a locking algorithm, and 3) trigger and interrupt staff for non-linear playback that samples from both the source and target staffs.   

	Melodrive is a web application that empowers users to generate original music pieces in seconds by following an intuitive workflow. The authors conducted a study that shows that an easier or more assisted process encourages non-musician users to make music. The system uses a patented music composition AI technology that acts as a virtual composer to the user. The AI generates new musical events given existing ones. Melodrive can be used in game environments since it can respond to affects by changing generative musical parameters such as harmony and rhythm. 
	Users can alter the instrumentation and musical styles of the generated music materials to improve their musical experience. These changes happen while the system attempts to maintain the basic musical content identified in the music.

	Melody Sauce by Evabeat\footnote{https://evabeat.com/} is a VST / Audio Unit FX plugin that generates melodies and hooks as MIDI. The plugin operates inside the user's DAW sequencer and supports melodic generation for electronic pop, dance, RnB and EDM styles. The user can quickly generate melodic ideas and assign them to project instruments while the song project is being played. The generative system uses melodic permutations and chance operations to generate melodic options. The users can control the key, tempo, the "mood" factor (light, dark or both), the note complexity of the generated melodies and harmony voicing options.

	FolkRNN\footnote{http://folkrnn.org} \cite{sturm2016music} is a web-based application for generating music that uses a ML model trained on transcriptions of folk music from Ireland and the UK from an online repository of work created by machines \footnote{https://themachinefolksession.org/}. The ML model consists of three hidden layers of long short-term memory (LSTM) units. The model is trained on over 23,000 crowd-sourced transcriptions. Many of the generated compositions have been performed, recorded and posted online.
	
	Jukedeck\footnote{https://www.jukedeck.com/} is a generative system using deep learning for assisted composition and music production. The system supports both audio and MIDI formats. The user can generate original music or select from a pre-generated library of audio tracks given the preferences regarding the music style, tempo and duration. The user can also edit the generated tracks.
	
	Amper Score™ by AmperMusic\footnote{https://www.ampermusic.com/} generates custom music in seconds for media such as video and podcast given a music style, length, and structure information (e.g. the timing of key moments). The system is built to be easy and intuitive for non-musicians. The generated content is rendered using a proprietary library of samples and purpose-built instruments.
	
	Spliqs\footnote{https://www.spliqs.com/} is an application for generating music using musical concepts called ``spliq". A spliq is an intelligent and adaptive entity that can generate music using a proprietary music representation and metacreative algorithms. The user selects a musical input such as the musical style or a specific song. He/she can use UI parameters to tweak or morph musical aspects (e.g. rhythm, melody) of the generated concept. The Spliqs platform is an ecosystem where spliqs reside. Spliqs can adapt to each other. Additionally, musical features of individual spliqs (e.g. rhythmic, harmonic, melodic) can be blended.

	
	Although these systems are useful for their intended purposes, neither do they offer the possibility for music generation based on a user corpus nor allow changing the underlying learning model. We argue that bringing these features will otherwise enable access to corpus-based style imitation with rich interactive control. This is the principal motivation of the Apollo project.
	
	\section{Apollo System Description}
	Apollo is a system for generating symbolic music using style imitation techniques and user corpora. The user can train style imitation models on a selected music corpus and subsequently generate musical phrases from the trained model. 
      
	The system is composed of three main features:
	\begin{itemize}
        \item The ability to prepare and manage a music corpus by uploading music files. The current system is limited to manipulating symbolic data encoded using the MIDI standard. The user can play, loop or streaming the MIDI content.
        \item The ability to select and train generative music models for style imitation purposes. The system currently hosts a constraint-based model and a MusicVAE model \cite{roberts2018hierarchical}. Apollo can dynamically discover and expose new python models to its GUI.
        \item The ability to generate symbolic music using the trained algorithm with control over parameters such as melodic typicality, harmonic following, number of measures and note density. These parameters help adjust the musical character of generated music materials. Those materials can then be exported for further editing or production activities.
    \end{itemize}
    
	Our work focuses on conventional western music with harmonic progressions, melodic content and rhythmic structures. The system runs as a desktop application. It can also be deployed as an application for the web browser. This can facilitate its accessibility to a number of communities of composers, researchers and producers. Finally, it can promote musical metacreative techniques to a broad community of non-coders or non-technical users. At this stage, we focus our prototype on the exploration and generation of musical phrases in the early stage of the compositional process. The generated phrases can later be saved or streamed into the user's native studio environment (e.g. Ableton Live) for further development.
    
	\subsection{Corpus Management}
\begin{figure}[h!]
        \centering
        \includegraphics[scale=.3]{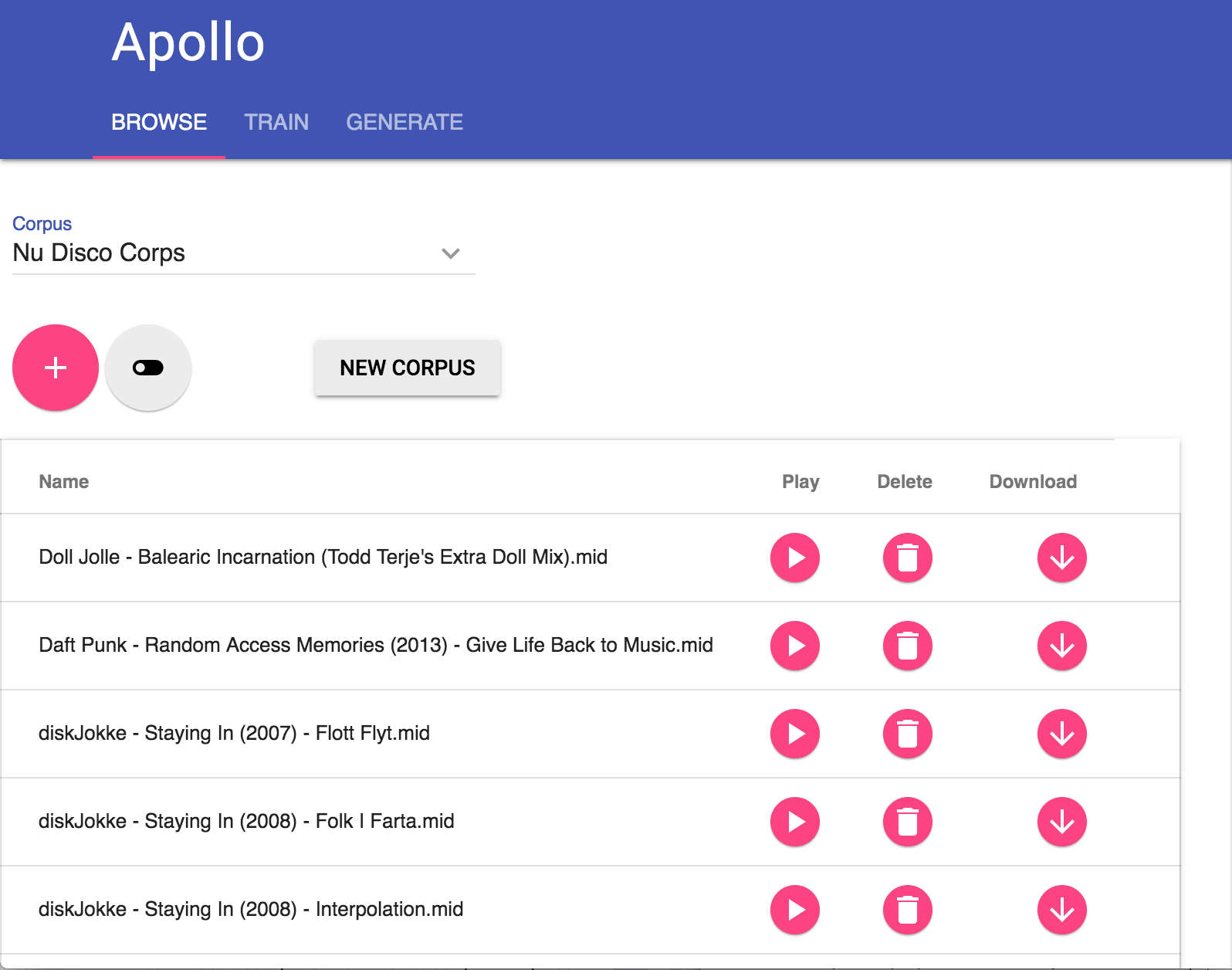}
        \caption{Browsing Interface}
        \label{fig:apollo browser}
    \end{figure}

    Apollo organizes music data through the creation, editing, and manipulation of musical corpora. The user first uploads a set of existing music files of MIDI format. This can be their own music catalogue. The catalogue can represent a particular musical style (e.g. artist, genre) or can be a more esoteric collection of music materials. It can also be musical phrases or fully-fledged musical compositions. The user can also work from pre-existing corpus available in the Apollo system. Figure \ref{fig:apollo browser} shows the user interface for corpus management. The user can create, save or load a music corpus. He/she can then edit it by adding or removing music pieces. The user can listen to the MIDI content of a music piece by playing it directly in Apollo or by streaming it to an external player via MIDI I/O ports.  This is achieved via a simple toggle button available in the browsing interface (Figure \ref{fig:apollo browser}).
	
	\subsection{Machine Training for Style Imitation}
\begin{figure}[h!]
        \centering
        \includegraphics[scale=.3]{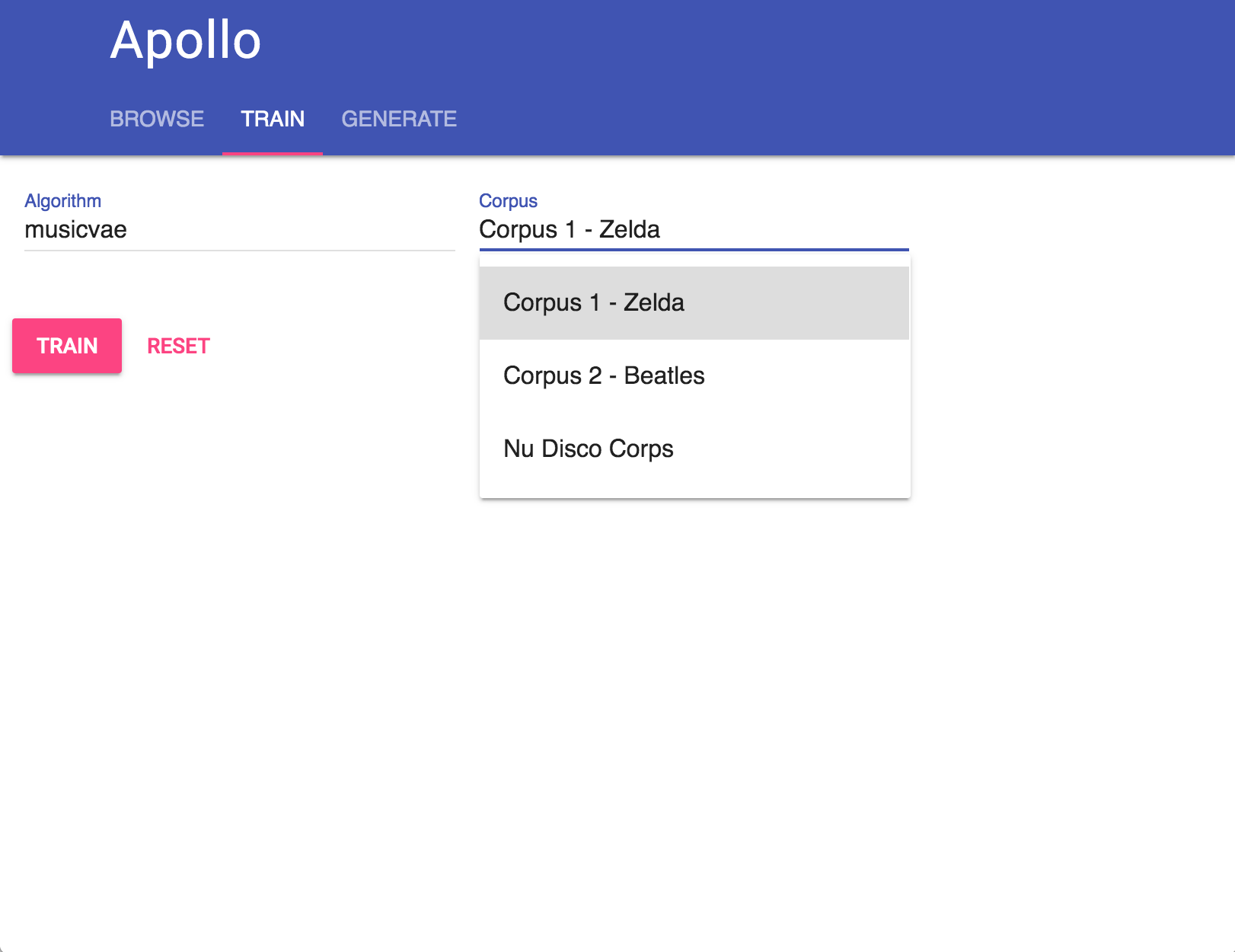}
        \caption{Training Interface}
        \label{fig:apollo training}
    \end{figure}

	The user can select the learning algorithm they wish to train.
    For each algorithm, a set of training parameters is discovered and available via the GUI. The user can adjust these parameters to achieve different training outcomes for the model. The user can then train the model. The music corpus is fed to the model according to the model training procedure. The training time depends on the algorithm performance and the size of the corpus. Currently, the Apollo System hosts our own model called \textit{Model1} which is a constraint-based style imitation algorithm, and the MusicVAE \cite{roberts2018hierarchical}. The Figure \ref{fig:apollo training} depicts the interface built for model training activities.
    
	 We developed Model1 in order to create a short feedback loop between training and generation. It can be trained in several minutes. The algorithm generates a trio of musical parts, including a bass, melody and drum part. The training process involves extracting statistical features from the MIDI files in the training corpus. The musical content is generated in two steps, as the rhythm structure is generated before specific pitches are selected. This decision was made out of necessity, as the search space was simply too large when rhythm and pitch were generated concurrently. For both steps, the constraint solver tries to find an optimal solution that exhibits stylistic characteristics of the corpus, while respecting preferences specified by the user via control parameters.
	 
	 The MusicVAE model is a variational autoencoder composed of a bidirectional encoder and hierarchical decoder to improve long-term music sequence learning. In the case of polyphonic training, the model is trained on "trio" MIDI files (drums, bass and melody channels) broken down into 16-bar musical sequences. The trained model we use in Apollo breaks down the user music corpus into 16-bar sequences and returns a set of latent vectors associated to that corpus.
	 We then use that set of latent vectors to generate new 16-bar musical sequences by randomly sampling from the set or by using the mean latent vector with some amount of Gaussian noise. 

	\subsection{Generating Music}
    The user can adjust the generation parameters. 
    This allows the user to influence the specific quality and variability of the generated musical phrases. The user can explore the generative space and behavior of the trained model and make decisions on if and how to edit the proposed music corpus. It is an opportunity to explore the generative process even after the music model has been trained.
    The user can then export generated musical phrases by streaming to their favorite DAW or saving them as MIDI files to their computer.
    
\begin{figure}[h!]
        \centering
        \includegraphics[scale=.3]{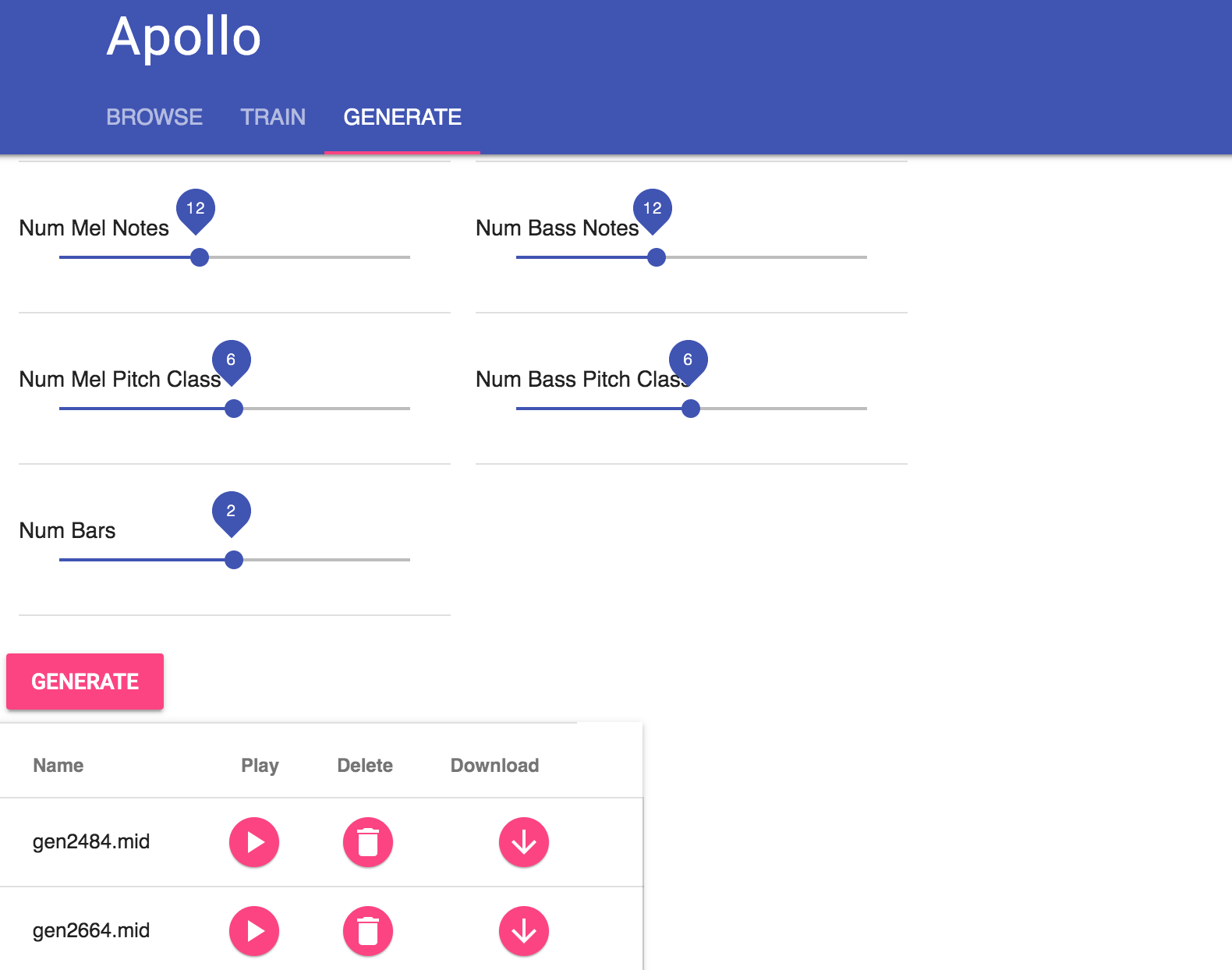}
        \caption{Generative Interface}
        \label{fig:apollo generate}
    \end{figure}

    The Figure \ref{fig:apollo generate} shows the user interface developed in the current Apollo prototype.
       
    \subsubsection{Generation Parameters}
     Apollo can discover model generation parameters and automatically render GUI controls for the user. To do this, we parse a JSON file that specifies the definition of those parameters. The specification model parameters is composed of a list of attributes for each parameters: a default value, description, display name (for the GUI), a maximum value, a minimum value, the programmatic name and the parameter type. An example from the MusicVAE model is:
    \begin{verbatim}
    { "default": 0.001,
    "desc": "amount of noise added to 
    latent vector",
    "display_name": "Noise Amount",
    "max": 1,
    "min": 0,
    "name": "noise_amount",
    "type": "float" }
    \end{verbatim}
    
    Some of the generation parameters that are exposed to the user interface for Model1 are the following:
    \begin{itemize}
        \item Melodic Typicality: This parameter controls the degree to which the generated melody adheres to the style learned by the trained model.
        \item Harmonic Following: The extent to which harmonies in the generated musical phrase follows each other coherently.
        \item Number of measures: The number of measures each generated musical phrase contains.
        \item Note Density: The average amount of notes found per measure in the generated musical phrases. It corresponds to the level of note saturation in a given phrase.
    \end{itemize}
    
    In the case of the MusicVAE, the generation parameters displayed are:
    \begin{itemize}
        \item Method: an integer value \{0,1,2\} which corresponds to the method of sampling a latent vector: 0 for random sampling in the set of latent vectors from the music corpus, 1 for sampling the mean vector, and 2 for randomly sampling from all latent vectors available to the MusicVAE. 
        \item Noise: controls the amount of Gaussian noise applied to the sampled latent vector.
        \item Temperature: is the softmax temperature of the MusicVAE which controls the overall randomness of the generated 16-bar musical sequences.
    \end{itemize}
    
    \subsection{Interactive Workflow}
\begin{figure}
        \centering
        \includegraphics[scale=.45]{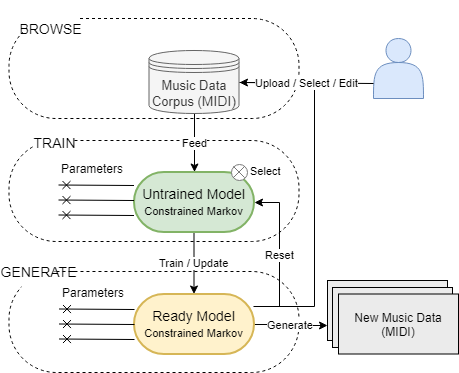}
        \caption{Interactive Workflow}
        \label{fig:apollo workflow}
    \end{figure}

    Figure \ref{fig:apollo workflow} summarizes the intended workflow based on the proposed features. The system breaks down into three interaction phases captured by three user interfaces mentioned in the previous sections: Corpus management (Browsing), Model training (Training) and Music Generation (Generation). The interactive process of music generation happens when the user is able to upload some music to the system, train and generate some outputs, listen to those outputs then adjust the music corpus, the training and generation parameters, and even the learning model to tune the expected generative behavior of the system. The user effectively uses the machine learning process as a way of composing music.
    
	\subsection{Implementation}
    The system is built as a Node.js web-based application using the MEAN stack (Mongodb, Express, Angular, Nodejs) as system architecture. We are using HTML5, CSS4 and Javascript for standard front-end development. We use Electron \footnote{https://electronjs.org}, an open-source framework for cross-platform desktop application development, to package and deploy our system. A demo of the Apollo system can be found at \textcolor{blue}{\textbf{http://metacreation.net/apollo/}}.
    
    We use Google's Material Design \footnote{https://material.io/} for the design of our user interface. Material is an open-source Design System that implements best practices for user interface design that includes principles of accessibility, responsive design, inclusiveness, cross-platforms and mobile-first. 
    
\begin{figure}[h!]
        \centering
        \includegraphics[scale=.45]{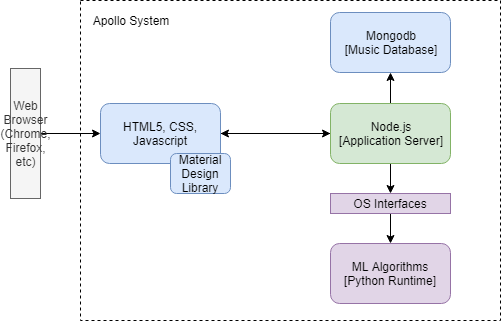}
        \caption{System Architecture}
        \label{fig:apollo architecture}
    \end{figure}

    A simple low-level file system interface to Node.js is used to provide interoperability between the application environment and the training models. New models are added to the \codeword{model_storage} folder so that they can be discovered by the Apollo system. The system currently only offers support for  Python models (as shown in Figure \ref{fig:apollo architecture}). 
    The Figure \ref{fig:apollo architecture} shows the deployment architecture for our system.
    
	\section{Discussion}
    
    Developing the Apollo system poses many interesting challenges such as manipulating a large set of binary files (for the MIDI music corpus) and transmitting them over the internet connection with limitations of HyperText Transfer Protocol (HTTP) protocols. This might require solutions such as those found in music cloud or streaming technologies such as Spotify or Last.fm.
	Despite the progress in real-time audio web technologies such as Web Audio API, there are still challenges to developing computer-assisted tools for the browser. For example, we encountered issues around MIDI playback \footnote{https://github.com/WebAudio/web-midi-api/issues/102}, latency during implementation and pending support of features by web browsers \footnote{https://bugs.chromium.org/p/chromium/issues/detail?id=471798} (Electron uses Chromium engine to run applications on desktop) and third party applications.
    
    The opportunity to manage more than one corpus enables for easy ways to create new corpora by manipulating existing ones. A good example is the ability to mix styles from two different corpora.
    this support can enable an environment where music researchers can test new algorithms on standardized music corpora which will be of significant value for the global research community. 
    
    The choice of Material as a user interface design framework assumes that best practices for web-based user interface design also extend to interfaces for music composition. This might not necessarily be true, especially in the context of a compositional process that follows the machine learning process. This is one of the assumptions that we will be evaluating in the future.
	
	\section{Future Work}
	
	There are many aspects of the Apollo project that will be worth investigating in the future.
	The first element of interest is to implement support for parsing of MIDI channel information and visual representations of MIDI tracks using graphical structures such as a piano roll. Access to MIDI channel information of a track can allow the user richer interactive generation such as for example, by training on partial MIDI information of the music corpus (say for MIDI channels of drums), the user can generate new musical phrases containing only drum pattern information, but still following the style of the corpus. Additionally, capturing basic music information and metadata such as tempo, key, track length on the user interface will be desirable.

    Another area for future work is user interface design. This area has shown interest to a community around designing intelligent musical interfaces for creativity-support in composition. Work in this area can enable us to better refine the Apollo software interfaces for better use. We intend to conduct preliminary user evaluation to understand the extent to which our system can creatively engage musicians in their artistic practice.
	
	\section{Conclusion}
	The Apollo system is a generative music environment that is built for use by a broad community of researchers, musicians  and artists. By enabling corpus management, model management and interactive machine learning, we make it possible for researchers to quickly test new models, and for musicians and artists to experiment with a selection of machine learning algorithms directly via a GUI. This represents an alternative design approach to other generative music systems allows us to continue to explore how such systems can be built to accommodate the activities of musicians, artists and researchers in computer-assisted composition.\\
	
\bibliographystyle{mume}
\bibliography{mume2019}
\end{document}